**High-quality, large-grain MoS$_2$ films grown on 100 mm sapphire substrates using a novel molybdenum precursor**

Paul Quayle[1], Bin Zhang[1], Jacob Leach[1], Brian Bersch[2], Joshua Robinson[2], and Shanee Pacley[3]

[1]Kyma Technologies Inc., 8829 Midway West Road, Raleigh, North Carolina 27617, USA

[2]Department of Materials Science and Engineering, Center for Two-Dimensional and Layered Materials, Pennsylvania State University, University Park, Pennsylvania 16802, United States

[3]Air Force Research Laboratory, Materials and Manufacturing Directorate, Nanoelectronics Branch, 3005 Hobson Way, WPAFB, OH 45433, USA

**Two-dimensional MoS$_2$ is a crystalline semiconductor with high potential for numerous technologies. Research in recent years has sought to exploit the direct band gap and high carrier mobility properties of monolayer MoS$_2$ for functional applications. To date, the production of MoS$_2$ has remained at the research level and samples are usually synthesized in small quantities using small yield, expensive techniques. In order to realize scalable MoS$_2$-based technology, large-area, high-quality and affordable MoS$_2$ wafers must become available. Here we report on MoS$_2$ films grown on 100 mm sapphire wafers by a chemical vapor deposition process utilizing hydrogen sulfide and molybdenum precursor mixtures consisting of Na$_2$MoO$_4$ and NaCl. The grains of these films are faceted, large-area, on the order of 1-75 μm in length on edge. Growth conditions are identified that yield monolayer MoS$_2$ films. Raman spectroscopy shows an E$_{2g}$ and A$_{1g}$ peak separation of 17.9-19.1 cm$^{-1}$. Photoluminescence spectroscopy shows that annealing the films post-growth suppresses trion emission. X-ray photoelectron spectroscopy of the annealed films shows an in increase in the S:Mo concentration ratio from 1.90:1.00 to 2.09:1.00.**

1. <u>Introduction</u>

Research is underway into a variety of applications based on transition metal dichalcogenides (TMDs) that include photonics [1] and optoelectronics [2], low power and high-speed nanoelectronics [3], electrocatalysts [4], and biosensors [5,6]. The monolayer TMD molybdenum disulphide (MoS$_2$) has received a great deal of interest in part because its large on /off current ratio [7], approximately 1.8 eV direct band gap, rich exciton structure, and strong light absorption and emission properties [8,9].

Studies of layered MoS$_2$ have been enabled by the availability of high quality samples produced from bulk molybdenite using the mechanical exfoliation technique. Samples of MoS$_2$ prepared in this way have shown field-effect mobilities as high as 120-200 cm$^2$V$^{-1}$s$^{-1}$ [10]. Layered MoS$_2$ can also be produced synthetically using various techniques [7], including metalorganic chemical vapor deposition and atomic layer deposition, which can produce wafer scale films [11,12]. The ability to grow $MoS_2$ by chemical vapor deposition (CVD) provides a path to scalability, however the electronic properties of CVD $MoS_2$ are consistently inferior to those of exfoliated $MoS_2$, with mobilities that are typically an order of magnitude lower than values for exfoliated $MoS_2$ [10]. The difference in apparent transport characteristics between CVD and exfoliated material has been attributed primarily to Coulomb scattering from subband trap states which perturb the true band mobility. Candidates for these localized charge states are sulphur vacancies, grain boundaries and dislocations, although charged interfacial states at contacts may also contribute [13]. Further deepening the understanding of the characteristics of CVD grown $MoS_2$, like the physical origins of subband traps, would potentially equip researchers to address such detrimental properties and achieve higher performance in devices.

In this paper, we summarize a brief product development study of $MoS_2$ films grown using a CVD method employing a Mo source composed of a mixture of sodium molybdate ($Na_2MoO_4$) and NaCl, and hydrogen sulfide ($H_2S$). Three samples are reported on and each consists predominantly of a monolayer film that covers nearly the entire surface of 100 mm wafers.

## 2. Experimental

### 2.1. Synthesis and Characterization Tools

Synthesis of the $MoS_2$ films was conducted in a horizontal furnace equipped with a 125 mm diameter quartz tube. A pair of removable tube liners was used to simplify reactor cleaning. Wafers were centered horizontally in the tube and held on a quartz wafer holder. Two alumina combustion boats (40w x 100l x 18h mm) were used to hold the Mo-precursor mixture. The combustion boats were positioned next to each other with their long sides in contact approximately 1-inch upstream from the wafer holder. Prior to each growth run, the combustion boats and wafer holders were cleaned in a sonicator using two-step IPA-acetone process and the sapphire substrates were cleaned using the standard RCA process and dried using $N_2$.

Raman and photoluminescence spectra were taken at room temperature in ambient atmosphere using a Horiba LabRam confocal system with 532 nm laser excitation. The laser was focused to a 1-2 μm spot size using a 50x objective. X-ray photoelectron spectroscopy was taken on a Surface Science Instruments M-Probe spectrometer with Al-Kα radiation. Measurements were done at $10^{-7}$ Pa with a scan resolution of 4 and a time step of 1 eV. The spot size was 1 mm.

## 2.2. Growth Parameters and Reaction Precursors

The growths were carried out at a substrate temperature of 820° C for 3 hours. The substrate was positioned horizontally, face down on a quartz substrate boat. The growth pressure was either 50 or 100 torr. The $N_2$ carrier gas flows was 5000 sccm and the $H_2S$ flow was either 5 sccm or 10 sccm. Each sample was grown with 36 grams of $Na_2MoO_4$, and a $Na_2MoO_4$:NaCl ratio of 2:1. A final annealing step was carried out for 1 hour at 820° C and 100 torr with gas flows of 20 sccm $H_2S$ and 5000 sccm $N_2$.

The Mo source used in this work is a two-component solid chemical mixture consisting of $Na_2MoO_4$ and NaCl. It has been observed that $Na_2MoO_4$ and NaCl react at elevated temperatures to form $Na_3ClMoO_4$, which then melts at 644 °C [14]. At elevated temperature, the mixture releases a gaseous Mo-precursor that allows for the synthesis of $MoS_2$. Additionally, we observed that the $Na_2MoO_4$ and NaCl mixture reacts strongly with the $H_2S$ process gas at the combustion boat.

An analysis of the composition of the Mo-precursor gas and the multivariate system is beyond the scope of this investigation. We speculate that the Mo-precursor gas consists in part of molybdenum dichloride dioxide ($MoO_2Cl_2$), which has been produced by the reaction of $MoO_3$ and $Cl_2$ [15]. In addition, various molybdenum chlorides ($MoCl_3$, $MoCl_4$, and $MoCl_5$) may contribute to the gaseous Mo-precursor flow [16,17]. The extent to which a gas phase Na compound is present is also unknown. The critical issue is that, as we discuss later, sodium is not present in the synthesized $MoS_2$ within measurement accuracy, according to XPS.

It is important to note that only scattered nucleation sites of $MoS_2$ form if NaCl is excluded from the Mo-source mixture.

Columns two and three of Table 1 provide the growth conditions for a series of experiments. Figure 1 shows a representative image of a $MoS_2$ film on a 100 mm sapphire substrate. Samples A, B, and C were grown under varying conditions to explore the influence of pressure and $H_2S$ flow. We observed that these parameters affect the morphology of the samples while the material quality, especially after annealing, remains mostly the same. We observed that 820 °C was the highest temperature that would yield $MoS_2$ films and crystallites that were uniform in texture and that the visual yellow color of the films was seemingly brighter at the high temperature. In general, the highest growth temperature possible is preferred for high crystalline quality materials.

| Sample ID | Pressure (torr) | Chalcogen Flow rate (sccm) | Small Crystallite Density (mm$^{-2}$) | Large Stack Density (mm$^{-2}$) | Monolayer Film Coverage (%) | $E_{2g}$ Raman Peak (cm$^{-1}$) | $A_{1g}$ Raman peak (cm$^{-1}$) | PL/$E_{2g}$ intensity ratio (as-grown) | PL/$E_{2g}$ intensity ratio (post-anneal) |
|---|---|---|---|---|---|---|---|---|---|
| A | 100 | 10 | 2090 | 240 | 73 | 383.5 | 401.4 | 4.43 | 2.08 |
| B | 100 | 5 | 43700* | 2 | 100 | 383.8 | 402.3 | 5.40 | 3.47 |
| C | 50 | 10 | 15450 | 930 | 100 | 382.6 | 401.7 | 3.18 | 2.43 |

**Table 1.** The varied growth parameters, film morphology characteristics, $E_{2g}$ and $A_{1g}$ Raman peak locations, and photoluminescence/Raman peak intensity ratios for Samples A, B, and C.

Footnote: * Sample B is predominantly bilayer at the front of the wafer. The listed value is for the bilayer crystallite density at the center of the wafer only.

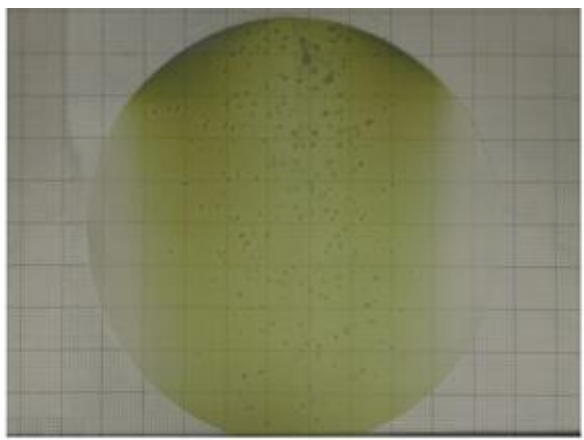

**Figure 1.** Representative image of Sample B. The major gridlines are spaced 1 cm. The dark speckling on the sample is due to growth deposits on the backside of the substrate, opposite to the MoS$_2$ film side.

### 3. Results and Discussion

#### 3.1. Sample Morphology

Synthesis of the monolayer MoS$_2$ proceeded from nucleation sites, which are scattered across the substrate surface. Bilayer and multilayer film growth also proceeded from the nucleation sites, and the relative rate of lateral versus vertical growth, along with the nucleation site density, determines the morphology of the sample.

The fourth column in Table 1 lists the small crystallite densities for the three samples. The small crystallites are defined as nucleation sites that continue to grow a few more layers, typically 2-4 layers, and their lateral size and density on the substrate vary with the growth parameters. An example of a small crystallite is marked in Fig. 2a.

The small crystallite densities of Samples A and C were counted from an optical microscope image at 20x magnification and the measurement areas were 310 μm x 450 μm. The small crystallite densities of Samples B were measured using a 50x objective, and the measurement areas spanned 130 μm x 175 μm. Measurements were taken at both the center of the wafer and near the front edge of the wafer, approximately 1.5 cm from the front edge. For Samples A and C, the values listed in Table 2 are an average of the measurements taken at the front and center of the wafers. Sample B is predominantly bilayer at the front of the wafer and the listed value is for the small crystallite density at the center of the wafer only.

The fifth column in Table 1 lists the large crystallite densities for the three samples. The large crystallites form at the nucleation sites when the vertical growth rate increases substantially, resulting in multilayer columnar structures. The large crystallite densities were measured in the same sample regions as the small crystallite densities. An example of a large crystallite is marked in Fig. 2a.

The sixth column in Table 1 lists the percentages of the substrate surface that are covered by monolayer films for the three samples. These percentages exclude the perimeter region of the substrate, which extends approximately 1 cm into the substrate. The percent coverages of the films across the measurement areas were determined by taking the ratio of image pixels associated with $MoS_2$ films to the entire image pixel count, using ImageJ software. This analysis was done based on the difference in contrast between the $MoS_2$ films and the bare sapphire. The measurement areas were the same as for the small and large crystallite densities.

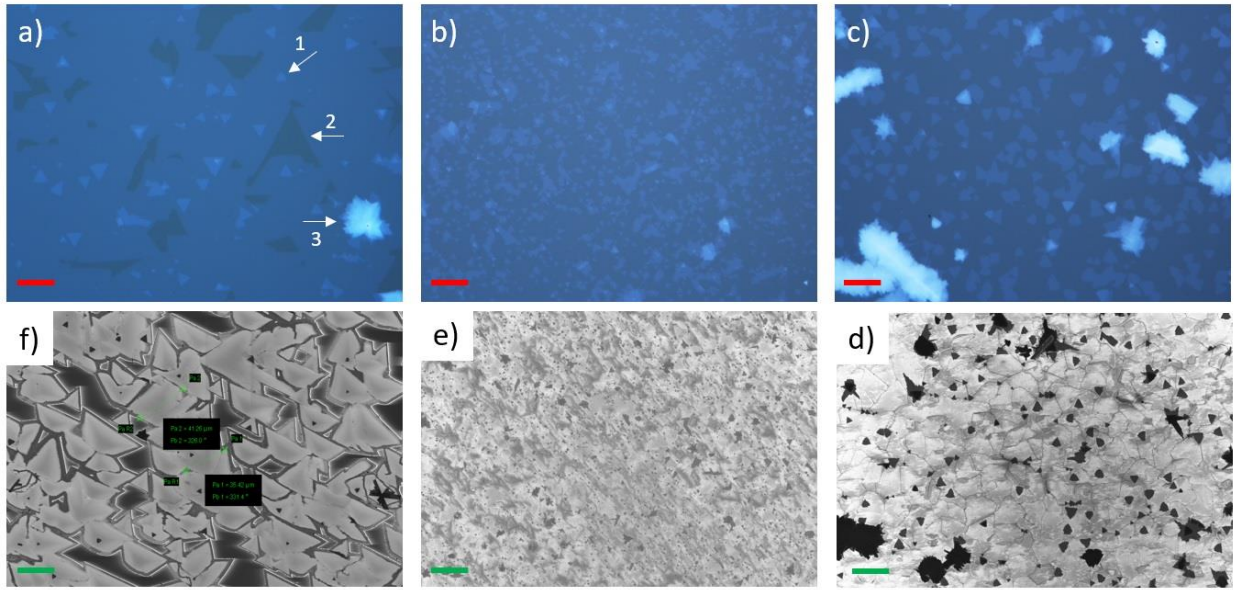

**Figure 2.** Optical microscope image of a) Sample A (a), b) Sample B, and c) Sample C. The scale bars in the OM images (red) represent 15 µm. The arrow labeled (1) in a) marks a small crystallite. The arrow labeled (2) in a) marks a bare sapphire region. The arrow labeled (3) in a) marks a large crystallite. Scanning electron microscope images of d) Sample C, d) Sample B, and f) Sample A. The scale bars in the SEM images (red) represent 25 µm. In the SEM images, the small and large crystallites appear black.

### 3.2. Growth Method

As stated in Section 2.2, the solid chemical Mo source reacts strongly with the $H_2S$ process gas, leading to the formation of parasitic deposition at the combustion boat. We observed that the amount of parasitic deposition at the combustion boat increases with the $H_2S$ composition in the process gases, which suggests that the partial pressures of $H_2S$ and the Mo-precursor gas are related; an increase in $H_2S$ partial pressure increases the consumption of the Mo-precursor at the combustion boat and inhibits the release of the gaseous Mo-precursor. This observation is reflected in the varying film morphologies.

#### 3.2.1. Sample A

Optical microscope and SEM images of Sample A and shown in Figs. 2a and 2f, respectively. This sample consists of relatively large grains which do not fully grow together into a continuous film. Edge lengths of the grains range from 30-75 µm. The large grain size and the discontinuous monolayer film coverage indicate that the nucleation density under the Sample A growth conditions is relatively low compared to Samples B and C.

### 3.2.2. Sample B

Optical microscope and SEM images of Sample B and shown in Figs. 2b and 2e, respectively. This sample has the smallest grain sizes, with edge lengths typically 1-5 µm. The grains in this film fully grew into each other yielding a continuous monolayer film. The small grain sizes of Sample B indicate that the nucleation density is relatively high. This sample also has the largest density of small crystallites.

### 3.2.3. Sample C

Optical microscope and SEM images of Sample C and shown in Figs. 2c and 2d, respectively. In this sample, the grain sizes are typically 20-30 µm on edge, slightly smaller than those of Sample A. The individual grains shown in Fig. 2d are highlighted by the darker grey perimeters, which are likely visible due to one-dimensional charge states that form at the interface of grains with different orientations [18]. The large crystallites described in Section 3.1 are visible as the black features in Fig. 2d.

### 3.2.4. Growth Analysis

Average grain sizes for Sample B are smaller than in Sample A but film does not cover the entire substrate surface in Sample A. This result indicates that the nucleation density is greater in Sample B than it is in Sample A. We conclude that the lower flow rate of $H_2S$ decreases the parasitic consumption of the Mo-precursor at the combustion boat, and thus allows for a greater amount of gaseous Mo-precursor to be released. As a result, there is a higher partial pressure of the Mo-precursor gas in the process flow during the growth of Sample B relative to Sample A, and thus a higher nucleation density, smaller grains, and continuous film coverage of the substrate, excluding substrate edge effects.

Sample C grew under the same gas flows as Sample A but at a lower pressure of 50 torr. These moderately high pressures are well within the diffusion limited evaporation regime. In this regime, the evaporation rate of a liquid is strongly affected by the mean free path of the gaseous atoms, and lower pressures result in higher evaporation rates. At 50 torr, the concentration of the Mo-precursor gas in the total process flow will be higher than at 100 torr, thus leading and the smaller grain sizes and higher nucleation density of Sample C relative to Sample A.

### 3.3. Optical Properties

The seventh and eighth columns in Table 1 list the measured Raman peak locations for the three samples. The $E_{2g}$ and $A_{1g}$ peak separations are 17.9, 18.5, and 19.1 cm$^{-1}$ for Samples A, B, and C, respectively, which are typical values for monolayer $MoS_2$ [19]. Figure 3a shows the Raman spectrum of Sample A, which is representative of the three samples.

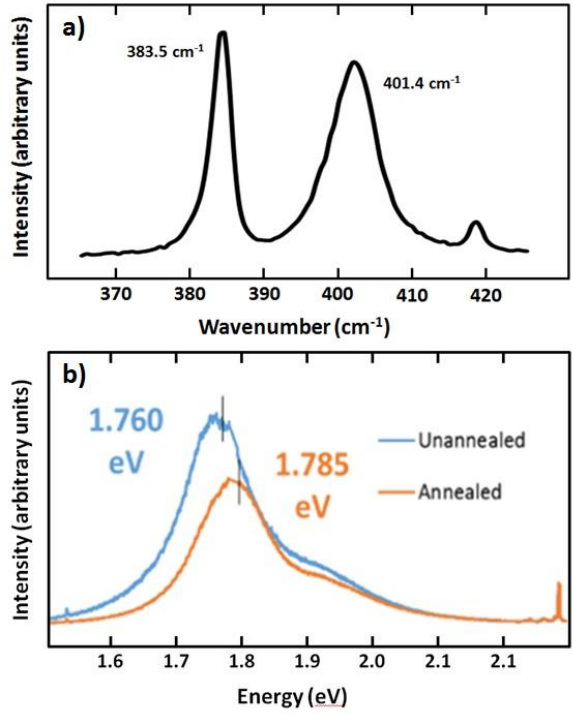

**Figure 3**. The (a) Raman and (b) photoluminescence spectra of Sample A.

The ninth and tenths columns list the PL/ $E_{2g}$ peak intensity ratio for the as-grown and post-annealed samples, respectively. The luminescence intensity ratio can be used to estimate the internal quantum efficiency of $MoS_2$ [20]. Here, we utilize the Raman peak intensity as an internal standard to measure the effect of annealing on the photoluminescence properties.

### 3.3.1. Oxygen Substitutional Defect Healing

All three samples – both annealed and unannealed – emit a strong PL signal with PL peak intensity/Raman $E_{2g}$ peak intensity ratios between 2-6. There is a substantial blue-shift in the PL peak position from 1.76 eV for as-grown samples to 1.79 eV for the annealed sample suggesting a shift in trionic character to more excitonic character (Fig. 2b) [21]. The trionic character of the $MoS_2$ PL signal usually indicates an increase in free electron density and n-type doping in samples which are often associated with oxygen bonding at defect sites [22,23]. In each case the PL/Raman intensity ratio decreases for annealed vs as-grown samples, which is likely due to an overall decrease in trion contribution to the total PL signal. Altogether, the observed blue-shifts in PL for the samples annealed in H$_2$S suggest a degree of oxygen substitutional defect healing in films.

3.4. Chemical Analysis

Figure 4 shows the chemical composition of an annealed portion of Sample B measured by XPS. The Mo peaks that are characteristic of MoS$_2$ are the Mo 3d +4 peaks at positions 229.8 and 232.9. The stoichiometry of this sample derived from these measurements is S:Mo = 2.09:1 [24]. This is an increase from a S:Mo = 1.90:1 measured on an unannealed portion of this sample. No signature sodium was measured in the samples.

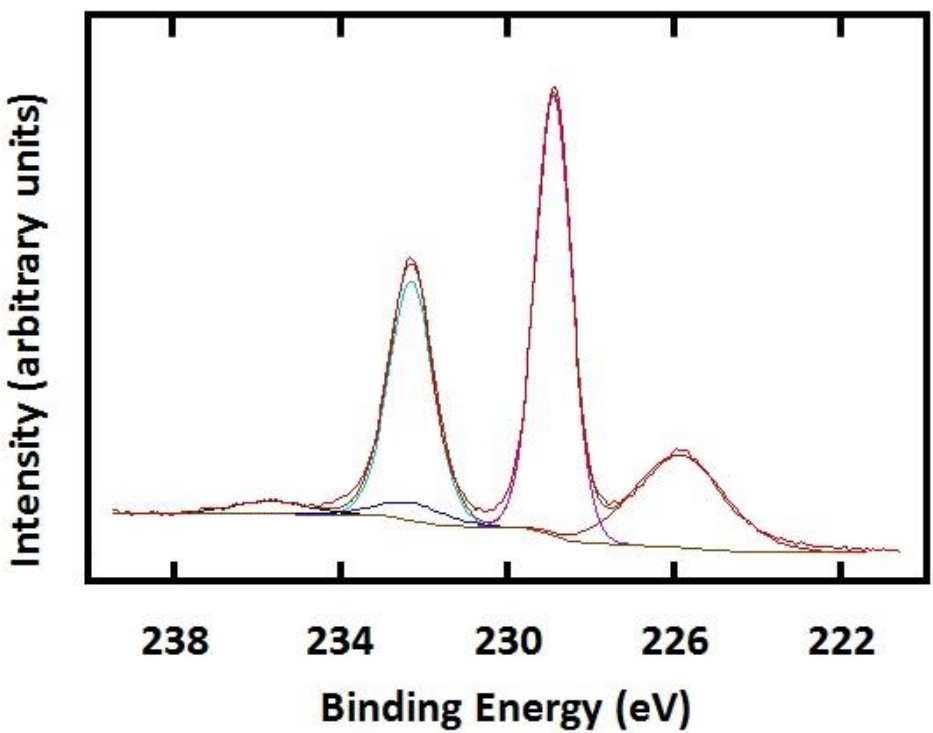

**Figure 4**. X-ray photoelectron spectrum of Sample B.

4. **Summary**

In summary, we have reported on monolayer MoS$_2$ grown by CVD on 100 mm sapphire substrates utilizing a Mo source consisting of Na$_2$MoO$_4$ and NaCl. Films grown using the novel precursor are chemically pure, exhibit strong photoluminescence that includes exciton structure, and are composed of faceted grains with edges length upwards of 75 μm. The inexpensive Mo-precursor identified here has demonstrated the potential to yield large-area, large-grain $MoS_2$ films. It also shows potential to yield high surface area, highly faceted multilayer stacks which could be of interest for catalytic applications.

Only three samples are reported on in this work, which is too small of a sample set to allow for a vigorous analysis of this growth method. However, the hope is that this study can inform the continuing development of $MoS_2$ film technology.

**Acknowledgements**

This work was supported by the Air Force Research Laboratory [FA 8650-11-D-5800] through Universal Technology Corporation. Special thanks to Ty Pollak at UTC.